# Long-Term Trend of Noctilucent Cloud Altitude Detected: A Comparison of Positional Data with Historical Measurements

Oleg S. Ugolnikov[a], Ilya S. Yankovsky[b,c], Maxim V. Klimenko[b], Egor O. Ugolnikov[d], Andrey M. Tatarnikov[d], Sergey G. Zheltoukhov[d]

*[a] Space Research Institute, Russian Academy of Sciences, Moscow, Russia*
*[b] Western Department of Pushkov Institute of Terrestrial Magnetism, Ionosphere, and Radio Wave Propagation, Russian Academy of Sciences, Kaliningrad, Russia*
*[c] Immanuel Kant Baltic Federal University, Kaliningrad, Russia*
*[d] Lomonosov Moscow State University, Moscow, Russia*

*Corresponding author e-mail: ougolnikov@gmail.com

**Abstract**
A large number of photometric and photographic position data of noctilucent clouds (NLC) obtained at 54-61°N, 29-50°E in 2020-2026 is used, comprising about 20,000 individual measurements of the altitude of NLC fragments recorded by a number of cameras. Comparison with published long-base triangulation measurements starting from 1890 reveals a descending trend in NLC altitude with a 99.5% confidence level. The current descent rate is −1.2 km per century, or half of the model prediction based on the current increase of $CO_2$ in the atmosphere. Other factors that can influence NLC altitude are considered: solar longitude (season), local solar time, lunar tides, latitude, and solar activity. Two independent methods of NLC altitude determination based on the same photometric data are considered and compared.

## 1. Introduction

Noctilucent clouds (NLC) were discovered in the late 19th century (Leslie, 1885; Backhouse, 1885) as bright-shining clouds appearing during summer twilight in the dusk segment. The fact that they are still illuminated by the Sun during this period pointed to their unusual height, more than 80 km. If the water-ice nature of these clouds is assumed (Wegener, 1912), extremely low temperatures (below 150 K) are required.

The first appearance of NLC followed the strong eruption of Krakatoa in 1883. The question of a possible relation between these events is still challenging. The clouds could appear as the result of an increase in methane ($CH_4$) and, subsequently, water vapor ($H_2O$) in the upper atmospheric layers (Thomas et al., 1989). Owing to the strong dependence of light-scattering intensity on particle size ($\sim r^6$), NLC brightness is very sensitive to the water vapor volume ratio.

NLC observation frequency tended to increase during the 20th century (Thomas and Olivero, 2001); however, this trend was not confirmed in some long-term reviews (Dalin et al., 2020). This may reflect an increase in $CH_4$ and $H_2O$, as well as the negative temperature trend in the middle and upper atmosphere caused by an increase in $CO_2$ (Roble and Dickinson, 1989). These relationships initiated the discussion about the possibility of using mesospheric ice as an indicator (the "Miner's Canary"; Thomas, 1996) of global climate change. The discussion was continued by von Zahn (2003) and Thomas et al. (2003).

Von Zahn (2003) argued that the NLC occurrence rate cannot be the basis of climate study, since it depends on the number of observers, weather conditions, and some other factors that can influence NLC properties: the season (or solar longitude) and solar activity. The absolute brightness of the cloud is difficult to compare, since the clouds were recorded by different detectors, mostly with no photometric calibration.

Moreover, the brightness and occurrence rate of NLC are driven by an increase in $H_2O$ and are not directly related to radiative cooling by $CO_2$: a decrease in mean temperature has been established over a wide range of altitudes, but the summer mesopause region is the exception (Berger and Lübken, 2011). Stratospheric and mesospheric cooling below the NLC layer produces the atmospheric shrinking effect (Roble and Dickinson, 1989). It can be observed as the long-term trend of mean NLC altitude. This was confirmed by calculations with the LIMA/MIMAS models (Lübken et al., 2018) with increasing and constant $CO_2$ and $H_2O$ in the atmosphere. Changing $CO_2$ caused a descending trend in the altitude of the mesospheric ice layer without changing the ice water content and occurrence rate, which were sensitive to the increase in $H_2O$.

In contrast to the occurrence rate, the mean altitude of NLC has been measured since the discovery of the clouds and can be compared. Von Zahn (2003) considers it the longest-measured parameter of noctilucent clouds. The same paper contains a brief review of available height measurement results, which varied in a narrow range from 82.2 to 83.6 km without a significant long-term trend. The variations were attributed to the different latitudes of observations; NLC altitude increased polewards.

However, one important point must be noted: the historical altitude measurements reviewed by von Zahn (2003) were conducted by triangulation from images taken at different observation sites, and referred to the brightest visual NLC fragments. Recent altitudes have been determined by lidars, which can detect weak and even invisible cloud particles. Owing to Mie scattering asymmetry, visual and photographic analysis of bright NLC in the dusk/dawn segment is very sensitive to the presence of the largest particles, while lidars can detect tiny particles that appear at higher altitudes. The height difference between the bottom of the ice layer, where particles reach their maximum size, and the cloud centroid altitude can reach 2 km (Gerding et al., 2013).

To perform the correct trend estimation, we must analyze results from similar measurement techniques. A review of triangulation heights from 1889 to 1979 (Gadsden and Schröder, 1989) also showed no significant trend, and additional data from Gadsden and Taylor (1994) did not change the picture. Long-term descending was difficult to detect owing to natural variations in NLC heights from 79 to 85 km.

The increase in $CO_2$ in the Earth's atmosphere is an accelerating process. It is represented in LIMA/MIMAS modeling (Lübken et al., 2018) for latitudes 55-61°N as a two-section line with different rates before and after 1960. The corresponding modeled rates of NLC altitude differ by a factor of 6.5. This effect can be more easily detected if a number of altitude measurements from recent times are added. If these measurements are conducted at different latitudes, seasons, lunar tide phases, and levels of solar activity, we can also reduce the effects of these factors on NLC altitude to find the long-term trend.

## 2. Historical data

The first estimation of the altitude of noctilucent clouds was made by Jesse (1885) from the principle that they must be illuminated by direct solar radiation. This led to an altitude above 80 km. A precise analysis by this technique was difficult owing to unknown properties of the extinction and refraction of tangent solar emission. The picture becomes rather complicated due to Chappuis absorption by stratospheric ozone, whose profile was not exactly known. However, in recent times, when the profiles of temperature, tropospheric and stratospheric aerosol, $O_3$ and $NO_2$ are known from satellite remote sensing, we can use the spectral or colorimetric data for an alternative estimation of NLC altitude (see below) basing on the same observations.

More exact NLC altitudes can be determined from triangulation data from two or more observation sites. The first measurements using this scheme were made in 1887 by V. Tserasky (Cerasky, 1890); the altitude was 79 km. However, there were two sources of large errors in this experiment. First, the accuracy of time synchronization $\Delta t$ between two observation sites was about several minutes. Taking into account that the velocity of NLC, $v$, can reach 50 m/s, this could bring an error $\Delta H \sim v\Delta t \cdot H/S$ that is more than 10 km in this case. Here $H$ is the altitude of the cloud, and $S$ is the baseline of Tserasky's experiment (the distance between the observation sites), 32.4 km.

The same magnitude of error can arise from a geometrical factor: the clouds were observed at low elevation above the horizon, $h$. For the case of low values of $S$ and $h$, we can write a simple expression for the geometrical error of altitude estimation, $\Delta H$. If the direction to the cloud is perpendicular to the baseline, the error is

$$\Delta H_{\perp} \sim \Delta\alpha\, H^2 / S\, h. \quad (1)$$

Here $\Delta\alpha$ is the error of cloud element sky position measurement. All angular values are expressed in radians. The error $\Delta H$ further increases if the baseline is directed towards the cloud; this is the case close to the geometry of Tserasky's observations:

$$\Delta H_{\parallel} \sim \Delta\alpha\, H^2 / S\, h^2. \quad (2)$$

Taking into account that the accuracy of angular measurements and the positions of the cloud, $h$, differ between observation sites, we can simply estimate $\Delta H$ by taking the highest error, $\Delta\alpha$, and the lowest elevation, $h$, among the observation sites. Since most of the historical measurements were made close to the horizon, and all types of errors are proportional to $S^{-1}$, we adopt the following criterion for including triangulation altitudes in our review and analysis: the baseline must be $S > H/2 = 40$ km, and exact time synchronization is necessary. The same criterion is applied to our recent measurements (see below), where $S$ is considered as the longest distance between the cameras included in the set, if their number is more than two.

The first long-base measurements to meet this criterion were performed by O. Jesse in 1890-1891 (Jesse, 1896) at several observation stations around Berlin, Germany, with a baseline of about 70 km. Jesse had provided the sky coordinates of the cloud fragments being analyzed, which allowed their latitude to be determined, as well as the statistical weight of each single measurement. We include these results with normalized weights ($w_i/w_S$), where $w_i$ is the weight provided by Jesse, and $w_S$ is the sum of the weights of the data during the twilight. The total weight of the data for each twilight is equal to unity. The same weighting system is applied to all historical data and to recent observations with the fixed camera set during the twilight. If historical data weights are not provided, they are set to be equal. If the sky coordinates of the cloud are not provided, its latitude is set to the latitude of the basic observation point plus 2°; this corresponds to the case of clouds at 20° above the northern horizon.

The next measurements were carried out by C. Störmer in 1932-1934 near Oslo, Norway (Störmer, 1933, 1935); the baseline in 1934 exceeded 100 km for the first time. Subsequent measurements with baselines longer than 40 km were conducted in 1958, 1965, and 1967 near Stockholm, Sweden (Grahn and Witt, 1971), and in 1959-1967 in the former USSR (Estonia and Latvia; Dirikis et al., 1966; Burov, 1966, 1967; Franzman and Franzman, 1967). The measurements up to that time are described in detail in a Russian review (Bronsten and Grishin, 1970).

During the last decades of the 20th century, triangulation measurements of NLC altitudes were performed in Scotland, UK (Taylor et al., 1984; Gadsden and Taylor, 1994). Measurements in 1986 had the longest baseline during the century, about 200 km. A review of altitude measurements until 1980 is presented in the book by Gadsden and Schröder (1989); no altitude trend was found until the end of the century. The possible increase in the rate of NLC depression predicted by Lübken et al. (2018) makes it possible to detect it now.

**3. Recent NLC altitude measurements**

A recent triangulation survey of NLC has been conducted since 2020 in the range 54-61°N, 29-50°E. About 20 locations and cameras were involved in different observation periods, mostly cameras of the Starvisor automatic net system. The cameras are divided into several groups with sizes of 50-200 km; the triangulation procedure is run for the cloud field recorded by a number of cameras of the same group. The total weight of the measurements from one camera set during twilight is also set to unity.

The procedure is based on projecting the observed cloud fields onto a surface at a fixed *a priori* altitude in the upper mesosphere using the ellipsoidal model of the Earth. This is followed by a fragmentation procedure, correlation analysis, and shift calculation for each fragment, which allows finding the exact altitude. Using the least-squares method makes it possible to involve more than two locations and cameras, increasing the accuracy and allowing the errors to be estimated correctly. Most of the cameras produce up to four frames per minute; analysis of the data allows the cloud fragment velocity to be found together with the altitude. Noctilucent clouds are analyzed in 2-3 minute intervals during the period of their visibility at the location of each camera group.

The method was initially described by Ugolnikov (2024) and applied to noctilucent clouds during the night of July 3-4, 2023, at 55-58°N, 34-38°E; these remain the brightest during the entire observation period described in this chapter. Ugolnikov et al. (2025, 2026a) expanded the procedure to a number of observations from recent years; the data from summer 2026 are added to the present research. The number of single measurements of NLC fragment altitude is about 20,000.

The basic RGB cameras of the first group (55.6°N, 36.6°E) have calibrated RAW format with known spectral bands, so their data can be used for photometry. This allowed the mean effective particle size to be found for the observed cloud field (Ugolnikov, 2023a) and its profile across the gravity wave pattern (Ugolnikov, 2023b). Spatially spread observations by this type of cameras made it possible to estimate the mean particle size for each fragment (Ugolnikov, 2024) and the nucleation/sublimation rate together with three velocity components (Ugolnikov et al., 2026b).

Three-color photometry can also be used to separate color effects caused by local atmospheric extinction, Mie scattering at different angles and by the immersion of NLC to the shadow of dense atmospheric layers. This provides an opportunity of an alternative estimation of NLC altitude using the same data. The technique can be considered an improvement of the umbral method used for the first altitude estimation right after the discovery of NLC. Here we analyze the change of NLC color ratios as the cloud immerses into the shadow of the stratospheric ozone (color turns blue) and then troposphere (color turns red). The accuracy of mean altitude can be as good as 200-300 meters if we know the profiles of temperature, aerosol extinction, $O_3$, and $NO_2$ in the troposphere and stratosphere (Ugolnikov, 2023a).

Since the method is applied to the same data, basically for bright clouds, agreement with triangulation is expected. However, a comparison for the same NLC ensembles in 2020-2024 (Ugolnikov et al., 2025) showed that photometric altitudes are on average about 0.5 km higher than triangulation altitudes. The authors suggested that photometric altitudes are overestimated due to the influence of multiple scattering (or scattering of the Earth's limb background) by NLC particles, which increases the NLC intensity near the Earth's umbra, especially in the blue part of the spectrum. However, a Monte Carlo simulation of radiative transfer (Ugolnikov, 2026) showed that this effect is negligibly small for noctilucent clouds.

The reason for this altitude shift has now been found: Ugolnikov (2023a) used a uniform model of the stratospheric aerosol particle size distribution. Aerosol extinction coefficients $E(H)$ for a given date, location, and altitude are taken from OMPS satellite data (NASA GES DISC, 2026) in four channels from 510 to 745 nm and then calculated for the RGB detectors of the cameras using the power law of wavelength dependence, $E \sim \lambda^{-n}$, where $n$ was taken as a constant value: $n = 2$. In this work, we calculate the value of $n$ from OMPS data as a function of altitude, $n(H)$. All other details of the procedure repeat those in Ugolnikov (2023a).

Figure 1 shows an example of a diagram of the color index (Red/Blue, effective wavelengths 578 and 469 nm, respectively) versus the local solar zenith angle as seen from the cloud fragment. The data for the brightest clouds in 2026, on the morning of July 18, are shown. The accuracy of each single color index measurement is low, especially near the Earth's umbra (solar zenith angle beyond 98°). However, the number of points is very high, which allows the mean altitude of the ensemble to be found by comparing the data with theoretical dependencies of color on solar zenith angle for different altitudes.

The triangulation technique allows the altitude of each fragment to be determined at a specific moment in time. Figure 2 shows the result of this procedure for 0h UT (local solar time about 2.5h) of the same twilight. We see that the mean triangulation altitude is even above the photometric altitude, but we have to note that this altitude value refers to a specific moment in time.

We compare the colorimetric altitudes with the averaged triangulation altitudes in the same twilights and camera systems. The diagram is shown in Figure 3. We see that the systematic shift of the altitudes has almost vanished. The colorimetric altitudes are now slightly lower than the triangulation data; the mean shift, taking errors into account, is −0.18 km. The basic analysis of the long-term trend is to be done in the following chapter based only on triangulation measurements. Colorimetric altitudes, the number of which is much smaller, are considered supplementary.

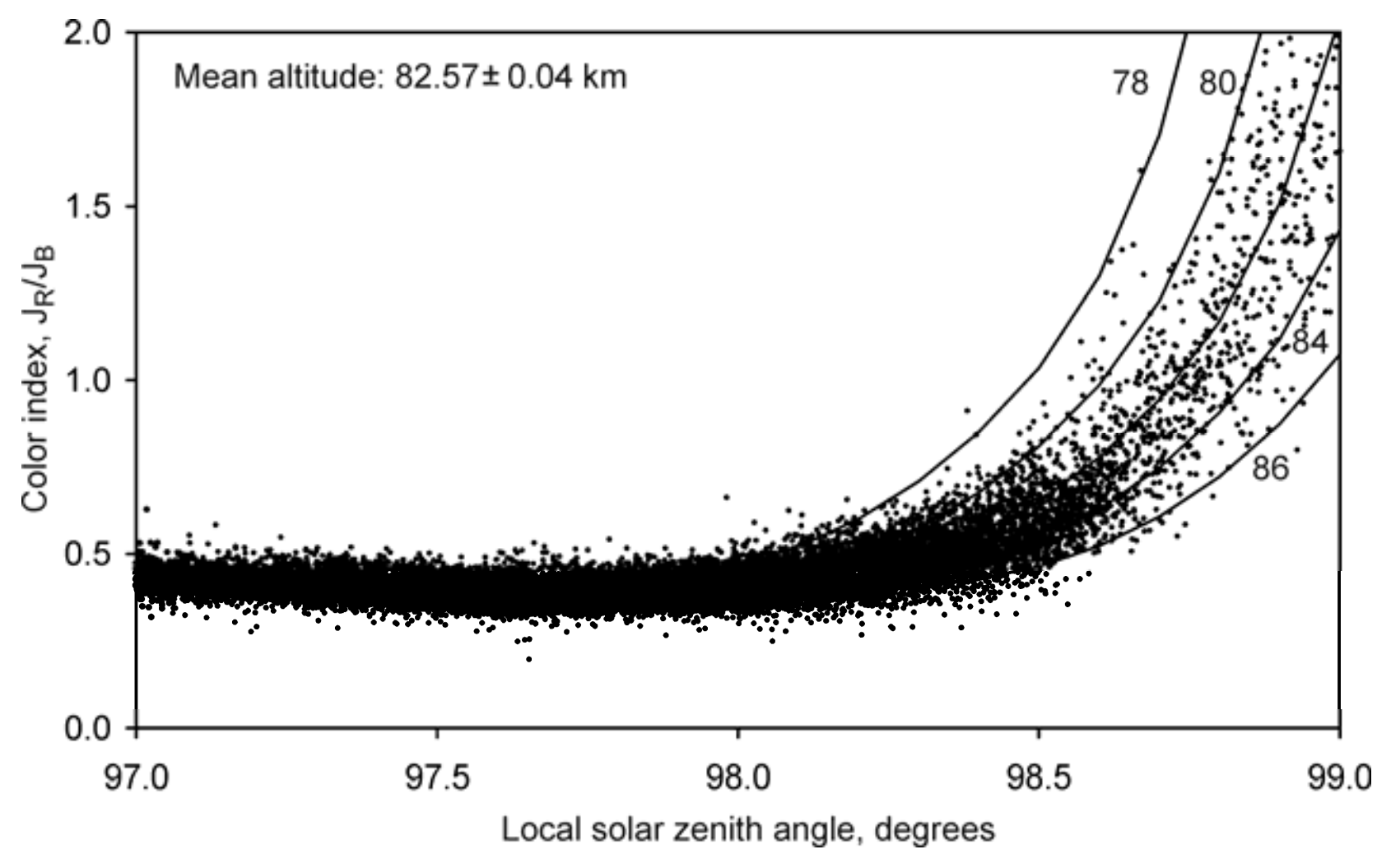

*Figure 1. Color index of noctilucent cloud fragments as a function of the solar zenith angle visible from the cloud, during the morning twilight of July 18, 2026, compared with radiative transfer modeling for different altitudes. The mean colorimetric altitude for the entire twilight period is shown.*

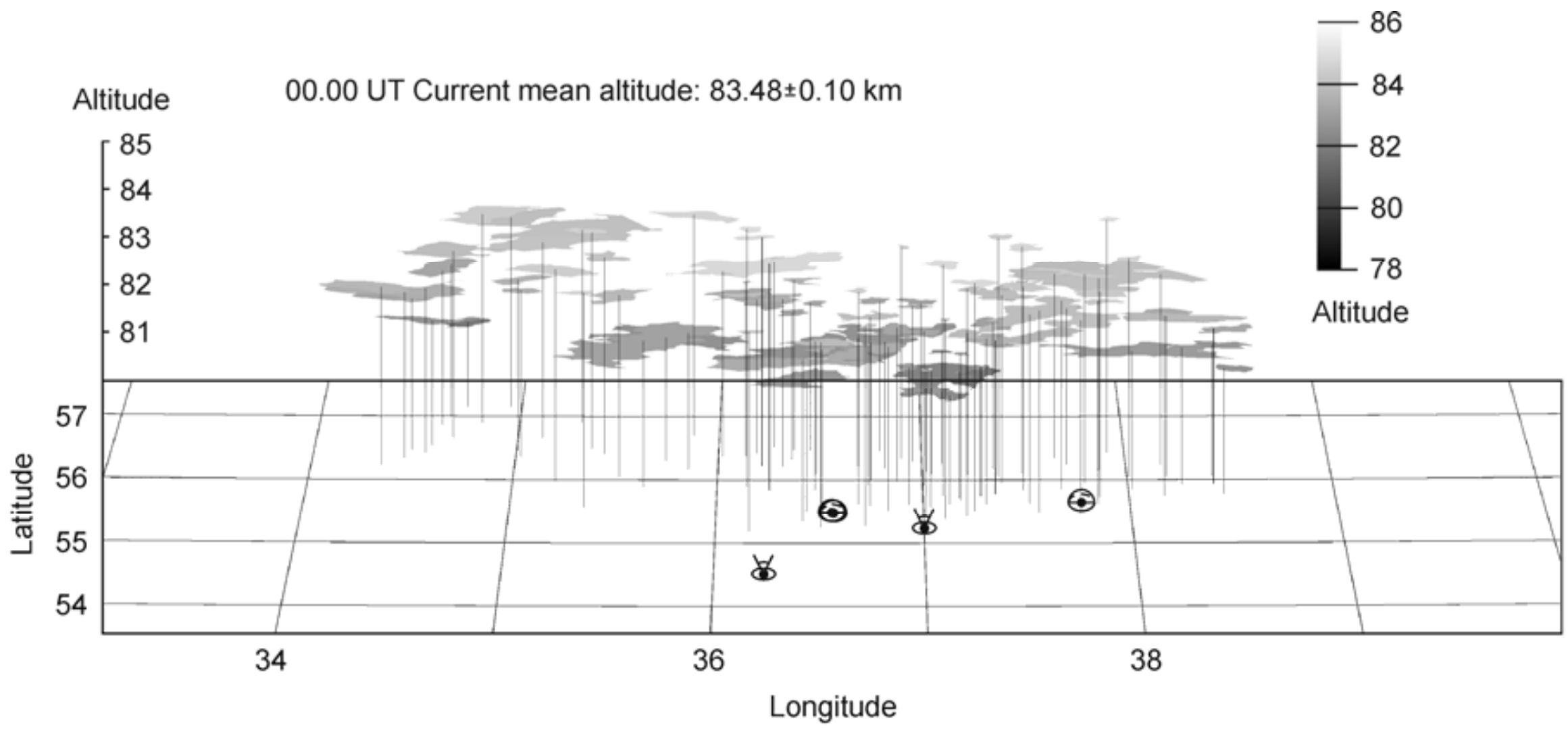

*Figure 2. Triangulation altitudes of NLC fragments, 00h00m UT, July 18, 2026. All-sky and northern sky cameras positions are shown.*

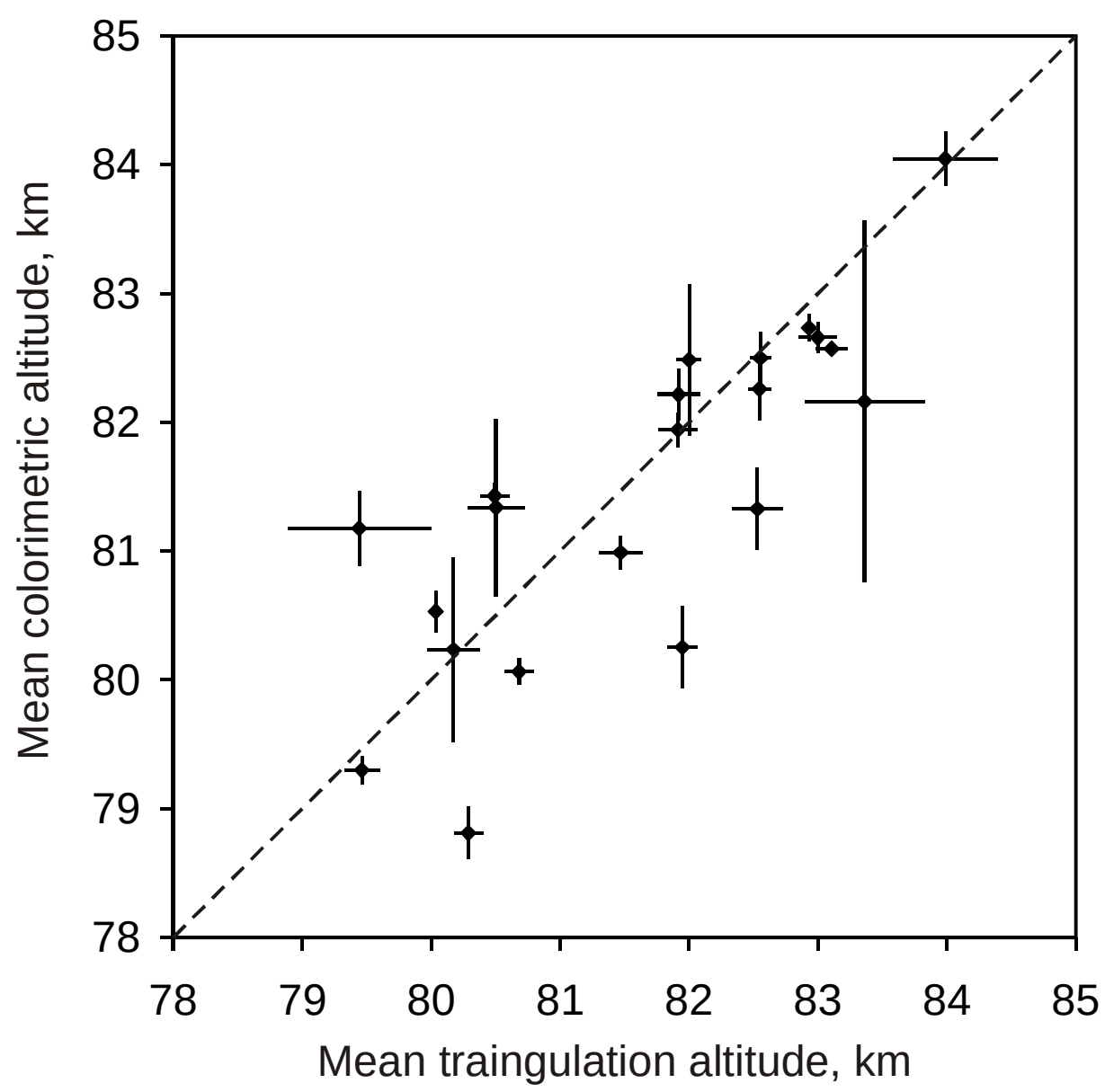

*Figure 3. Comparison of mean triangulation and colorimetric altitudes averaged over the twilights from 2020 to 2026.*

## 4. List of factors included in the model

To determine the long-term trend in the altitude of noctilucent clouds correctly, we must take into account all factors those can influence it. We describe each one and how it is represented in the following least-squares procedure.

### *4.1. Long-term trend*

This is the basic factor studied in this work, possibly related to the increase of $CO_2$ in the stratosphere and mesosphere and to the "atmospheric shrinking" effect. It is expected to be non-linear. In the LIMA/MIMAS model (Lübken et al., 2018), it is defined in the simplest form as a two-segment broken line with a kink point at year 1960. The model predicts the long-term change of mean NLC altitude expressed in kilometers as a function of time *t* expressed in years:

$$H = H_0 - T(t);$$
$$T(t) = 0.0036\ (t - 1880),\ \ t \leq 1960;$$
$$T(t) = 0.288 + 0.0233\ (t - 1960),\ \ t > 1960. \qquad (3)$$

The value of $H_0$ corresponding to the mean NLC altitude for 1880 is 83.8 km; however, it is expected to be lower for bright visual clouds. We consider $H_0$ as the first free parameter of the least-squares procedure. The second dimensionless free parameter, *K*, characterizes the ratio of the real to the modeled long-term trend of NLC altitude:

$$H = H_0 - K{\cdot}T(t) \qquad (4)$$

Here, *K*=1 means that the trend is in agreement with the model calculations by LIMA/MIMAS, while *K*=0 refers to the case in which no trend exists. *K* is the basic parameter to be found in this paper.

*4.2. Seasonal dependence*

It is natural that the difference in physical conditions in the mesosphere at the beginning, middle, and end of the NLC summer season can lead to a seasonal dependence of NLC altitude. The visible clouds consist of large particles that form at the bottom of the supersaturated layer below the mesopause. This layer reaches its maximum thickness after the solstice; the minimum of altitude is expected at this time.

A decrease in NLC altitude near the solstice was detected by ALOMAR lidar (Fiedler et al., 2003). A subsequent study (Lübken et al., 2008) showed that the altitude reaches its minimum in July, several weeks after the solstice, together with the maximum in NLC occurrence rate. The seasonal minimum of cloud altitude after the solstice was confirmed by SOFIE satellite measurements (Li et al., 2024) and was seen in our triangulation data (Ugolnikov et al., 2025, 2026a).

The seasonal dependence of NLC altitude is now modeled as a second-degree polynomial:

$$\Delta H_S = A\ (l_S - \pi/2) + B\ (l_S - \pi/2)^2 \qquad (5)$$

Here $l_S$ is the ecliptic longitude of the Sun in radians. If we express the value $(l_S - \pi/2)$ in degrees, it will be close to the time after the solstice in days. The parameters *A* and *B* are to be found in our analysis.

*4.3. Diurnal variation*

The properties of noctilucent clouds can experience significant diurnal variations, as was initially described by Jensen et al. (1989). Von Zahn et al. (1998) found variations of NLC altitude by lidar measurements, which were also confirmed by Fiedler et al. (2005). During the time interval around local midnight, the mean altitude increases; this is basically part of the semidiurnal tidal cycle (Klostermeyer, 2001).

Since the image measurements are restricted to a short period around midnight, the variations of altitude can be modeled in linear form:

$$\Delta H_D = D\ \tau \qquad (6)$$

Here $\tau$ is the local solar time expressed in hours. It is negative before midnight.

*4.4. Lunar tide phase*

The dependence of NLC occurrence rate on the phase of the lunar tide was noticed by Kropotkina and Shefov (1975). The lunar tide is probably responsible for 12-hour (von Savigny et al., 2017), half-monthly, and monthly (Dalin et al, 2006) variations in NLC occurrence.

The amplitude of lunar tides increases near the new moon and full moon, when the lunar and solar tides add to each other. In this case, the tides reach a maximum near local noon and local midnight, when NLC are observed. The effect of lunar tides on NLC altitude was estimated to be insignificant. To take it into account in the simplest form, we add the term

$$\Delta H_T = L \cos 2(l_L - l_S) \qquad (7)$$

The cosine of the double difference of the lunar and solar ecliptic longitudes reaches unity at new moon and full moon, decreasing to −1 at the first and last quarters. The value of $L$ represents the sensitivity of NLC altitude to lunar tides.

*4.5. Meridional profile*

The dependence of mean NLC heights on latitude has been a topic of study since the late 20th century; however, Thomas and Olivero (1986) had not found a significant effect. Von Zahn (2003) reported a positive relation of about 0.03 km per degree; however, as we discussed above, high-latitude recent lidar altitudes were compared with low-latitude historical triangulation data. The same value was found in the lidar investigation of Lübken et al. (2008).

In our study, the latitude range of measurements is not wide. We use a linear relation for the altitude correction:

$$\Delta H_M = M\,(\varphi - \varphi_0) \qquad (8)$$

Here $\varphi$ is the latitude of the cloud fragment in degrees, and $\varphi_0$ is the typical mean latitude of NLC fragments in the anlalysis. The value $\varphi_0 = 58°$ is chosen. We note that the choice of $\varphi_0$ does not change the estimation of parameter $M$ in the linear least-squares procedure but allows the errors of this procedure to be reduced.

*4.6. Solar activity*

Solar activity significantly influences the properties of noctilucent clouds, basically through variations of the solar Ly-$\alpha$ flux reaching the upper mesosphere. During solar maximum epochs, the mesosphere is heated while the volume ratio of $H_2O$ decreases; both effects lead to a decrease of ice mass in the mesosphere (Hervig and Siskind, 2006).

The relation with mean NLC altitude is more doubtful. Recent LIMA/MIMAS modeling (Vellalassery et al., 2024) shows that the mean altitude can increase by about 0.3 km during the solar maximum epoch. This effect is barely seen in lidar data (Fiedler et al., 2009). Our recent NLC triangulation altitudes include epochs of solar minimum and solar maximum. This allows the effect of solar activity to be traced as

$$\Delta H_C = C\,(W - W_0) \qquad (9)$$

Here $W$ is the Wolf number averaged over a period of ±2 days relative to the moment of NLC observation. Wolf numbers are taken from the SILSO database (Clette and Lefèvre, 2015), where data from the very beginning of NLC measurements in 19th century are available. Analogously to the previous chapter, we take $W_0 = 90$ as the characteristic mean Wolf number.

## 5. Results

Combining all the factors listed above, we build the basic equation for the altitude of noctilucent clouds:

$$H = H_0 - K\cdot T(t) + A\,(l_S - \pi/2) + B\,(l_S - \pi/2)^2 + \\ + D\,\tau + L\cos 2(l_L - l_S) + M\,(\varphi - \varphi_0) + C\,(W - W_0) \quad (10)$$

This equation is written for each single measurement of NLC fragment altitude from 1890 to 2026, with the weighting system described above. For recent measurements since 2020, only triangulation altitudes are initially involved. The system is solved by the least-squares method. The results are presented in Table 1.

We see that the last four parameters (dependencies on local solar time, lunar tide phase, latitude, and Wolf number) are below the errors of their determinations; however, the values of the NLC height gradients with respect to local solar time and Wolf number are in good agreement with the estimates in the papers listed above. The only significant factors are the long-term trend and the seasonal dependence.

The value of the long-term trend of noctilucent cloud altitude is 2.9 times higher than its error, which means that it is detected at better than the 99.5% confidence level. However, it is two times lower than the LIMA/MIMAS prediction by Lübken et al. (2018). This corresponds to an descending rate of about –1.2 km per century during the present epoch.

Figure 4 shows the altitude values averaged over the data of one camera group during the twilight. The statistical summary made above can be confirmed by simple calculations. We can take the mean altitude of all measurements before 2000 (82.52 ± 0.16 km) and the mean value from 2020s triangulation (81.83 ± 0.18 km). The difference is 0.69 ± 0.24 km, which corresponds to almost the same confidence level (99.5%). This value is about half of the decrease of NLC altitude predicted by Lübken et al. (2018) during the period of its fast change since 1960.

The value of $K$ slightly increases (0.58 ± 0.18) if we add the colorimetric altitudes from the all-sky RGB cameras described above to our analysis with the same weight as the triangulation altitudes for the same camera group and twilight. Here we must note that the amount of colorimetric data on NLC altitude is less than the triangulation data (see Figure 4).

| Description | | Value |
|---|---|---|
| | $H_0$, km | 82.8 ± 0.3 |
| Trend, related to model | $K$ | 0.52 ± 0.18 |
| Seasonal dependence | $A$, km | –3.5 ± 0.9 |
| | $B$, km | 5.9 ± 1.2 |
| Diurnal variations | $D$, km/hr | 0.05 ± 0.08 |
| Lunar tides response | $L$, km | 0.00 ± 0.19 |
| Meridional profile | $M$, km/deg | 0.01 ± 0.06 |
| Solar activity response | $C$, km | 0.0013 ± 0.0019 |

*Table 1. Parameters defining the altitude of NLC, results of least-square procedure.*

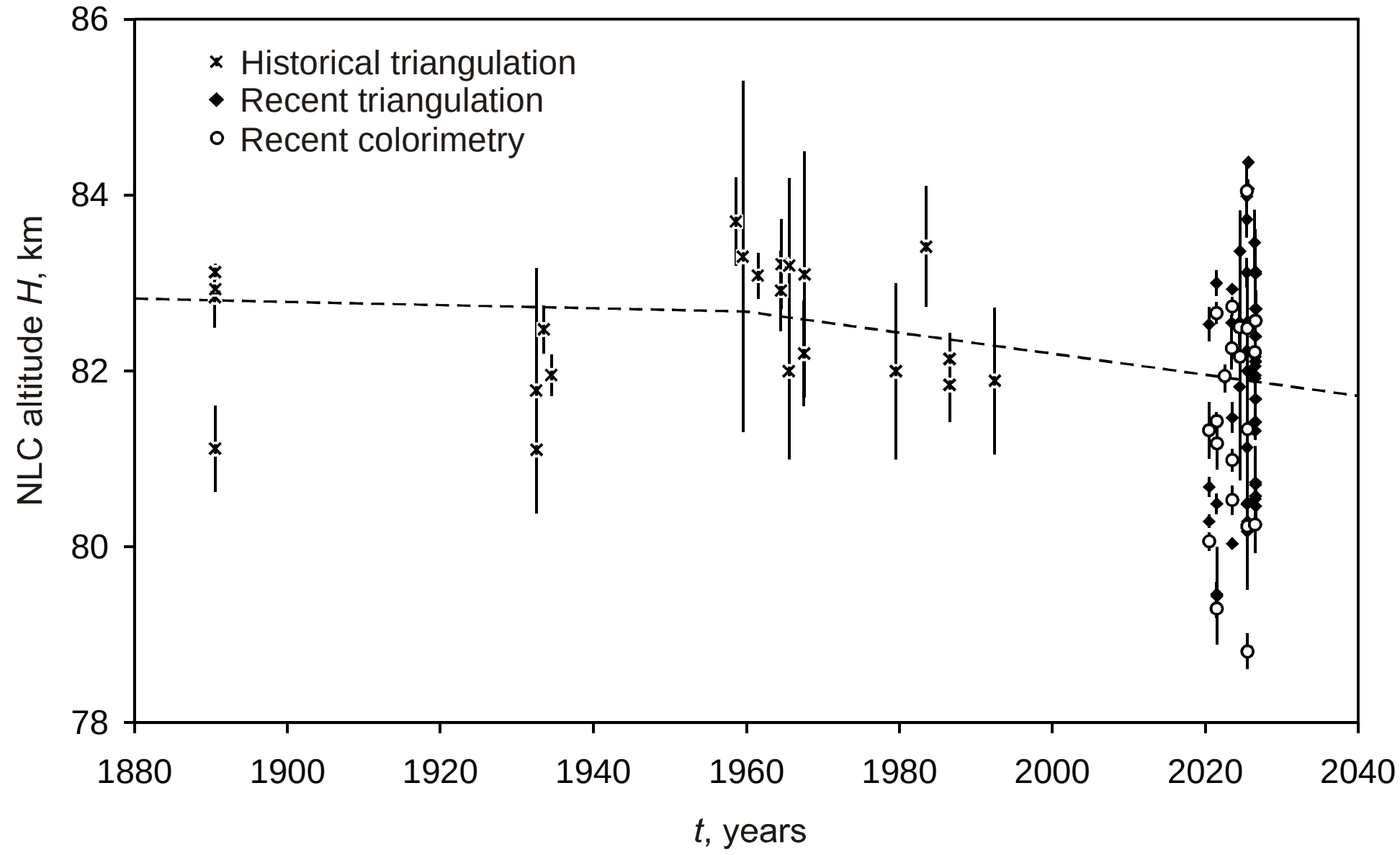

*Figure 4. The values of NLC altitudes averaged over the entire twilights. The dashed line corresponds to the LIMA/MIMAS trend multiplied by K = 0.52.*

Another significant effect seen in the NLC altitude data is the seasonal variation. Figure 5 shows a diagram of solar ecliptic longitude (or date relative to the solstice) versus NLC height. We see that the minimum of mean altitude is reached near July 10, in agreement with previous works listed in Chapter 4.2.

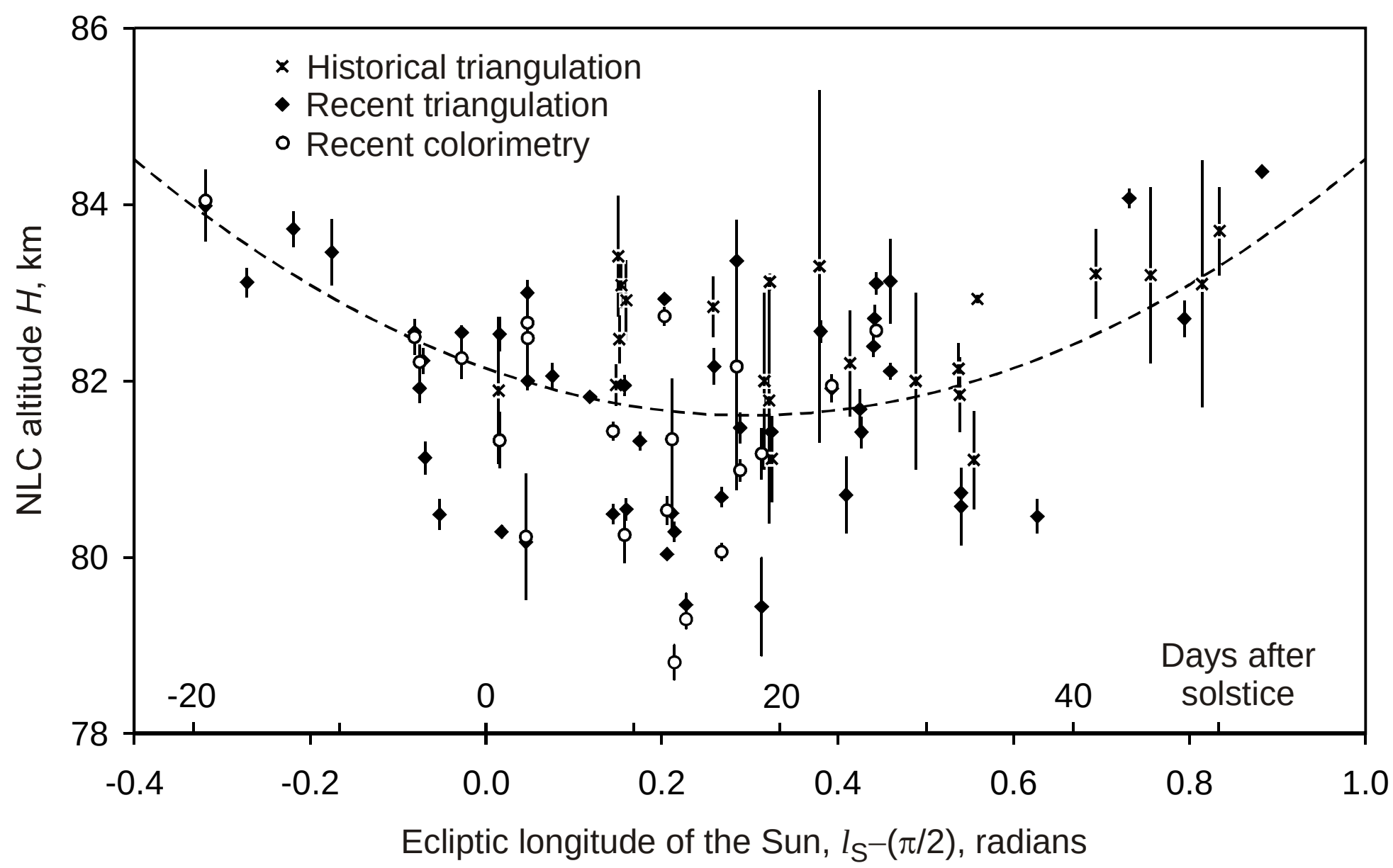

*Figure 5. Seasonal dependence of NLC altitudes; the same data as in Figure 4. The dashed line corresponds to approximation (5) for triangulation altitudes.*

## 6. Discussion and conclusion

The long-term trend of the altitude of noctilucent clouds has been a subject of investigation using triangulation data over a time interval exceeding 100 years, but this was not successful during the 20th century. The question was important due to its relation to the "shrinking effect" caused by the increase of $CO_2$ in the atmosphere. Models show that the rate of NLC descent must increase since the late 20th century, as $CO_2$ does. This fact, together with the opportunity of using big data technologies and the significant increase in the accuracy of photometry and position measurements, makes the question of NLC long-term trend detection relevant at the present time.

The basic difficulties are the natural scatter of NLC altitudes, exceeding the value of their mean depression over almost 150 years of observation, and the systematic shift of altitudes measured by different techniques, for example, lidars and photo imaging. A good example is the 1-km difference between the pre-industrial mean altitude referred to backscattering in LIMA/MIMAS models and the value of $H_0$ found here based on historical measurements. This determined the choice of triangulation as the basic technique for trend study. It is supplemented by colorimetric measurements of altitude using the same cloud data, which systematic shift was reduced to –0.18 km after the correction of the procedure made here. However, the basic conclusion is made based on position analysis only.

A comparison of the measurements during the late 19th and 20th centuries with recent altitudes during the last years showed a difference of about 0.7 km at a confidence level better than 99%. Applying these results to the model, we see that the real trend is half of the predicted value, the current NLC descent rate is 1.2 ± 0.4 km per century. As we can see in Figures 4 and 5, the upper limit of the NLC altitude distribution remains nearly constant during the season and on the long timescale. The change of the lower altitude is about twice the change of the mean altitude, being close to the descending rate predicted by Lübken et al. (2018).

Further triangulation measurements of NLC altitude in the near future can increase the accuracy of the mean current altitude value and detect the response to other factors that have been deemed insignificant in this paper. However, the accuracy of the trend value will remain limited by the low number and precision of historical measurements. Significant progress is expected when long-term arrays of lidar, satellite, and imaging data during the 21st century become available.

## Acknowledgments

Authors are grateful to Daria Kotova (University of Oslo) for her help with data collection, and to Sergey Potanin (Moscow State University), Nikolay Pertsev and Vladimir Perminov (Institute of Atmospheric Physics, Russian Academy of Sciences), Dmitry Aleshin, Natalia Andrievskaya, Dmitry Belov, Dmirty Gerasimov, Olga Golubeva, Viktor Kazatykh, Stanislav Korotkiy, Anatoly Morozov, Sergey Nagornov, Alexey Popov, Alexander Shadov, Yuri Stepanychev, Alexey Sushkov, Ekaterina Tipikina, Egor Volkov, and Damir Yalyshev for maintaining the noctilucent cloud imaging system used in this paper**.**

## References

Backhouse, T.W., 1885. The luminous cirrus cloud of June and July. Meteorological Magazine, 20, 133.

Berger, U., Lübken, F.-J., 2011. Mesospheric temperature trends at mid-latitudes in summer. Geophys. Res. Lett. 38, L22804. https://doi.org/10.1029/2011GL049528

Bronshten, V.A., Grishin, N.I., 1970. Noctilucent clouds. Nauka, Moscow (in Russian).

Burov, M.I., 1966. Determination of parameters of noctilucent clouds in 1964, Meteorologicheskiye Issled Serebristye Oblaka (Meteorolog. Inv. Noct. Clouds), II, 12, 33-46 (in Russian)

Burov, M.I., 1967. Spatial and dynamic characteristics of the noctilucent clouds. Symposium “Noctilucent cloud”, Tallinn, 1966, 200-207.

Ceraski, W. (Tseraskii, V.), 1890. Sur les nuages lumineux, Annales de l'Observatoire de Moscou, II, 177-180.

Clette, F., Lefevre, L., 2015. SILSO Sunspot Number V2.0. Published by WDC SILSO - Royal Observatory of Belgium (ROB). https://doi.org/10.24414/qnza-ac80

Dalin, P., Pertsev, N.N., Romeiko, V.A., 2006. Significance of lunar impact on noctilucent clouds. J. Atm. Sol-Terr, Phys. 68, 1653-1663. https://doi.org/10.1016/j.jastp.2006.06.005

Dalin, P., Perminov, V., Pertsev, N., Romejko, V., 2020. Updated long-term trends in mesopause temperature, airglow emissions, and noctilucent clouds. J. Geophys. Res. Atmos. 125, e2019JD030814. https://doi.org/10.1029/2019JD030814

Dirikis, M.A., Evdokimenko, S.V., Franzman, Yu.L., 1966. Determination of spatial coordinates of noctilucent clouds. Meteorologicheskiye Issled Serebristye Oblaka (Meteorolog. Inv. Noct. Clouds), II, 12, 47-51 (in Russian)

Fiedler, J., Baumgarten, G., von Cossart, G. 2003. Noctilucent clouds above ALOMAR between 1997 and 2001: Occurrence and properties. J. Geophys. Res. 108, D88453. https://doi.org/10.1029/2002JD002419

Fiedler, J., Baumgarten, G., von Cossart, G. 2005. Mean diurnal variations of noctilucent clouds during 7 years of lidar observations at ALOMAR. Ann. Geophys. 23, 1175-1181. https://doi.org/10.5194/angeo-23-1175-2005

Fiedler, J., Baumgarten, G., Lübken, F.-J., 2009. NLC observations during one solar cycle above ALOMAR. J. Atm. Sol-Terr, Phys. 71, 424–433. https://doi.org/10.1016/j.jastp.2008.11.010

Franzman, S.V., Franzman Yu.L., 1967. Calculation of the coordinates of noctilucent clouds on BESM-2. “Observation of noctilucent clouds”, Nauka, 123-135 (in Russian)

Gadsden, M., Schröder, W., 1989. Nocilucent Clouds. Springer-Verlag, Berlin.

Gadsden, M., Taylor, M.J., 1994. Measurements of noctilucent cloud heights: a bench mark for changes in the mesosphere. J. Atm. Sol-Terr, Phys. 56, 461-466. https://doi.org/10.1016/0021-9169(94)90194-5

Gerding, M.. Kopp, M., Hoffmann, P., Höffner, J., Lübken, F.-J., 2013. Diurnal variations of midlatitude NLC parameters observed by daylight-capable lidar and their relation to ambient parameters. Geophys. Res. Lett. 40, 6390-6394. https://doi.org/10.1002/2013GL057955

Grahn, S., Witt, G., 1971. Repp AP-4 Inst Meteorol Univ. Stockholm.

Hervig, M., Siskind, D., 2006. Decadal and inter-hemispheric variability in polar mesospheric clouds, water vapor, and temperature. J. Atm. Sol-Terr, Phys. 68, 30–41. https://doi.org/10.1016/j.jastp.2005.08.010

Jensen, E., Thomas, G.E., Toon, O.B., 1989. On the Diurnal Variation of Noctilucent Clouds. J. Geophys. Res. 94, 14693-14702. https://doi.org/10.1029/JD094iD12p14693

Jesse, O., 1885. Die auffallenden abenderscheinungen am himmel. Meteorol. Ztschr 2 (20), 311-312.

Jesse, O., 1896. Die höhe der leuchtenden Nachtwolken. Astron. Nachrichten. 140, № 3347, 161-168.

Klostermeyer, J., 2001. Effect of tidal variability on the mean diurnal variation of noctilucent clouds. J. Geophys. Res. 106, D9, 9749-975. https://doi.org/10.1029/2000JD900838

Kropotkina, E.P., Shefov, N.N., 1975. Influence of lunar tides on the probability of noctilucent clouds appearance. Atmospheric and Oceanic Physics 11, 11, 184-185 (in Russian).

Leslie, R.C., 1885. Sky glows. Nature. 32, 245.

Li, Y., Gao, H., Sun, Sh., Li, X., 2024. Correlation between peak height of polar mesospheric clouds and mesopause temperature. Atmosphere. 15, 1149. https://doi.org/10.3390/atmos15101149

Lübken, F.-J., Baumgarten, G., Fiedler, J., Gerding, M., Höffner, J., Berger, U., 2008. Seasonal and latitudinal variation of noctilucent cloud altitudes. Geophys. Res. Lett. 35, L06801, http://dx.doi.org/10.1029/2007GL032281

Lübken, F.-J., Berger, U., Baumgarten, G., 2018. On the anthropogenic impact on long-term evolution of noctilucent clouds. On the anthropogenic impact on long-term evolution of noctilucent clouds. Geophys. Res. Lett. v.45, p.6681-6689. https://doi.org/10.1029/2018GL077719

NASA Goddard Earth science data and information services center (GES DISC). Accessed Sep 22, 2026, at: https://omps.gesdisc.eosdis.nasa.gov/data/SNPP_OMPS_Level2/ OMPS_NPP_LP_L2_AER_DAILY.2.5/2026/

Roble, R.G., Dickinson, R.E., 1989. How will changes in carbon dioxide and methane modify the mean structure of the mesosphere and thermosphere? Geophys. Res. Lett. 16. 1441-1444. https://doi.org/10.1029/GL016i012p01441

Störmer, C., 1933. Height and velocity of luminous night clouds observed in Norway, 1932. Oslo Univ. Obs. Publ. 6.

Störmer, C., 1935. Measurements of luminous night clouds in Norway, 1933 and 1934. Astrophysica Norwegica 1, 3, 87-114.

Taylor, M.J., Hapgood, M.A., Simmons, D.A.R., 1984. The effect of atmospheric screening of the visible border of noctilucent clouds. J. Atm. Sol-Terr. Phys. 46, 363-372. https://doi.org/10.1016/0021-9169(84)90121-1

Thomas, G.E., Olivero, J.J., 1986. The heights of polar mesospheric clouds. Geophys. Res. Lett. 13, 1403-1406. https://doi.org/10.1029/GL013i013p01403

Thomas, G.E., Olivero, J.J., Jensen, E.J., Schröder W., Toon, O.B., 1989. Relation Between Increasing Methane and the Presence of Ice Clouds at the Mesopause, Nature, 338, 490-492. https://doi.org/10.1038/338490a0

Thomas, G.E., 1996. Is the polar mesosphere the miner's canary of global change? Adv. Space Res. 18, (3)149-158. https://doi.org/10.1016/0273-1177(95)00855-9

Thomas, G.E., Olivero, J.J., 2001. Noctilucent clouds as possible indicators of global change in the mesosphere. Adv. Space Res. 28, 7, 937-946. https://doi.org/10.1016/S0273-1177(01)80021-1

Thomas, G.E., Olivero, J.J., DeLand, M., Shettle, E.P.. 2003. Comment on "Are noctilucent clouds truly a 'Miner's Canary' for Global Change?". Eos, Trans. Amer. Geophys. Union 84(36), 352-353. https://doi.org/10.1029/2003EO360008

Ugolnikov, O.S., 2023a. Altitude and particle size measurements of noctilucent clouds by RGB photometry: Radiative transfer and correlation analysis. J. Quant. Spectrosc. Radiat. Transf. v.296, p.108433. https://doi.org/10.1016/j.jqsrt.2022.108433

Ugolnikov, O.S., 2023b. Cross-wave profiles of altitude and particle size of noctilucent clouds in the case of one-dimensional small-scale gravity wave pattern. J. Atm. Sol. Terr. Phys. 243, 106024. https://doi.org/10.1016/j.jastp.2023.106024

Ugolnikov, O.S., 2024. Noctilucent clouds altitude and particle size mapping based on spread observations by ground-based all-sky cameras. J. Atm. Sol. Terr. Phys. v.259, p.106242. https://doi.org/10.1016/j.jastp.2024.106242

Ugolnikov, O.S., Pertsev, N.N., Perminov, V.I., et al., 2025. Five-years altitude statistics of noctilucent clouds based on multi-site wide-field camera survey. J. Atm. Sol. Terr. Phys. v.269, p.106491. https://doi.org/10.1016/j.jastp.2025.106491

Ugolnikov, O.S., 2026. Multiple scattering effects in noctilucent clouds: Numerical estimation and application to altitude and particle size measurements. J. Quant. Spectrosc. Radiat. Transf. 353, 109836. https://doi.org/10.1016/j.jqsrt.2026.109836

Ugolnikov, O.S., Yankovsky, I.S., Pertsev, N.N., et al., 2026a. Noctilucent clouds modulated by strong 5-day planetary wave in 2025: amplitudes, phases and altitudes based on ground-based observations and satellite temperature data. Adv. Space Res. 77, 12466-12475. https://doi.org/10.1016/j.asr.2026.04.038

Ugolnikov, O.S., Golubeva, O.Yu., Ugolnikov, E.O., 2026b. Three-dimensional velocity analysis and particle size dynamics from multi-site RGB-photometry of noctilucent clouds. J. Atm. Sol-Terr. Phys. 287, 106969. https://doi.org/10.1016/j.jastp.2026.106969

Vellalassery, A., Baumgarten, G., Grygalashvyly, M., Lübken, F.-J., 2024. Long-term evolution in noctilucent clouds' response to the solar cycle: a model-based study. Atmosphere 15, 88-103. https://doi.org/10.3390/atmos15010088

Von Savigny, C., DeLand, M.T., Schwartz, M.J., 2017. First identification of lunar tides in satellite observations of noctilucent clouds. J. Atm. Sol. Terr. Phys. 162, 116-121. https://doi.org/10.1016/j.jastp.2016.07.002

Von Zahn, U., von Cossart, G., Fielder, J., Rees, D., 1998. Tidal variations of noctilucent clouds measured at 69°N latitude by groundbased lidar. Geophys. Res. Lett. 25, 9, 1289-1292. https://doi.org/10.1029/98GL00546

Von Zahn, U., 2003. Are noctilucent clouds a "miner's canary" for global change? Eos, Trans. Amer. Geophys. Union, 84(28), 261-264. https://doi.org/10.1029/2003EO280001

Wegener A., 1912. Die Erforschung der obersten Atmospharenschichten. Gerl. Beitrage Geophys. 11, 102.